\documentclass[conference]{IEEEtran}
\makeatletter
\def\ps@headings{%
\def\@oddhead{\mbox{}\scriptsize\rightmark \hfil \thepage}%
\def\@evenhead{\scriptsize\thepage \hfil \leftmark\mbox{}}%
\def\@oddfoot{}%
\def\@evenfoot{}}
\makeatother
\usepackage[pdftex]{graphicx}
\usepackage{snapshot}
\usepackage{amssymb}
\usepackage{mathabx}
\usepackage[ruled,vlined,linesnumbered]{algorithm2e}
\usepackage{paralist}

\newtheorem{theorem}{Theorem}[section]
\newtheorem{property}[theorem]{Property}

\ifCLASSINFOpdf
\else
\fi
\begin{document}

\title{FlowME: Lattice-based Traffic Measurement}

\author{\IEEEauthorblockN{Petko Valtchev \IEEEauthorrefmark{1}, Omar Mounaouar \IEEEauthorrefmark{1}, Omar Cherkaoui \IEEEauthorrefmark{1}, Alexandar  Dimitrov \IEEEauthorrefmark{1} and  Laurent Marchand \IEEEauthorrefmark{2}}\IEEEauthorblockA{\IEEEauthorrefmark{1}Department d'informatique, UQAM, Montreal, \\Email: petko.valtchev@uqam.ca} \IEEEauthorblockA{\IEEEauthorrefmark{2} Ericsson Montreal, Canada 
}}

\maketitle

\begin{abstract}
%
%
%
%
Flow-based traffic measurement is a very challenging problem: Managing counters for each individual traffic flow in hardware resources knowingly struggle to scale with high-speed links.
In this paper we propose a novel lattice theory-based approach that 
improves flow-based measurement performances and scales by keeping the number of the maintained hardware counters to a minimum (result mathematically established in the paper). The crucial contribution of the lattice is to map the computational semantics of the packet processing to user requests for traffic measurement thus allowing for a better-informed and focused counter assignment. 
An implementation over an Openflow switch, FlowME, was developed and evaluated upon its memory usage, performance overhead, and processing effort to generate the minimal solution. Experimental results indicate a significant decrease in resource consumption.


%

\end{abstract}

\IEEEpeerreviewmaketitle

\section{Introduction}

Network traffic measurement is an essential activity that allows network managers to get the visibility required for daily operations and network evolution planning. 
Tools to observe per-flow traffic must scale with a wide spectrum of applications, flows and queries while maintaining the performance of the underlying hardware, achieving accurate traffic measurements and operating at wire speed~\cite{Estan:2002:NDT:964725.633056}. Conventional solutions like NetFlow sample traffic and send per-flow statistics to a remote server to exploit in user applications, thus incurring inaccurate statistics and intensive resource and network bandwidth usage. 
%
Recent works~\cite{Yuan:2011:PTP:1959441.1959451,GHANNADIAN2009} in application-aware traffic measurement use also prior knowledge about users requirements, i.e. user queries, to achieve adaptive measurements but they require dedicated packet classification mechanisms to carry out the measurement task. 
\begin{figure}[b]
     \centering
     \includegraphics[width=.7\linewidth]{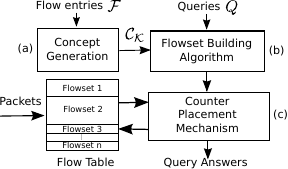}
	 \caption{Architecture Overview}
     \label{fig:AO}
\end{figure}

A significant drawback of current solutions is they largely ignore the computational structure of the packet processing and the induced query-to-flow associations. 
We regard these associations as crucial and believe that only a structure that correctly expresses them has the richness and flexibility to support the search for an optimal counter assignment. Thus, we turn to concept lattices and formal concept analysis  (FCA)~\cite{fca:ganter-99}.
%

Our lattice-based traffic measurement method, FlowME, enables fine-grain querying of the network traffic flows and extracting of the query-bound flow measurements. Figure~\ref{fig:AO} illustrates its main components: 
As a main supporting structure, a hierarchy of high-level, flows-to-matchfields abstractions, the \textit{concepts}, is constructed (a) and each query is mapped to a unique node thereof, its \textit{target} concept (comprising the  answer set of flows). The hierarchy, or the \textit{concept lattice}, factors out commonalities in the answer sets by turning them into concepts. Targets induce a sub-hierarchy where each node - called \textit{projection} - corresponds to an intersection of answer sets. In order to avoid the redundancies in the resulting family of sets, each flow is mapped to a minimal projection, its \textit{ground} (b). We show that by assigning a counter to every projection, a system of counters is obtained which is both minimal in size and allows all the queries to get a precise answer (c).
%
%
As a result, in FlowME the hardware counters are kept for disjoint sets of flow entries (instead of passively monitoring all flows). Moreover, the memory usage is further reduced by focusing only on flows matching user queries.

	The contributions of this work are as follows.
	\begin{itemize}
	\item We propose four algorithms for constructing/maintaining the concept lattice and its projection substructure (section \ref{sec:algorithms}). The projection algorithms are original methods that underlie the central task of partitioning the global set of flows into disjoint subsets (to be assigned a counter each).

	\item We prove that the number of counters established in this way is minimal w.r.t. the requirements of: 
	     $(i)$ answering all active queries, and $(ii)$ assigning a single counter to a flow (Theorems~\ref{thm:assign-corr} and~\ref{thm:assign-min} in Section~\ref{sec:proofs}).
	\item The FlowME solution can be used for flow-based measurement in a wide range of network devices and is expected, in  particular, to enable effective monitoring in Openflow switches.  Its implementation over an Openflow Pizzabox switch largely outperforms a per-flow counter assignment at a reasonable computational cost.
	
 \end{itemize}

In the remainder of the paper, we first present our lattice construction and updating algorithms (section~\ref{sec:algorithms}). In section~\ref{sec:proofs} the mathematical foundations of our solution are summarized and its efficiency/minimality are proven. Section~\ref{sec:results} presents the major components of our implementation as well as its performance evaluation results. We discuss related work in Section~\ref{sec:related-work} and conclude in Section~\ref{sec:conclusion}.

%
%
%
%

\section{Algorithms}
\label{sec:algorithms}
%
%
%
%
%

\begin{figure}[htb]
     \centering
     \includegraphics[width=.8\linewidth]{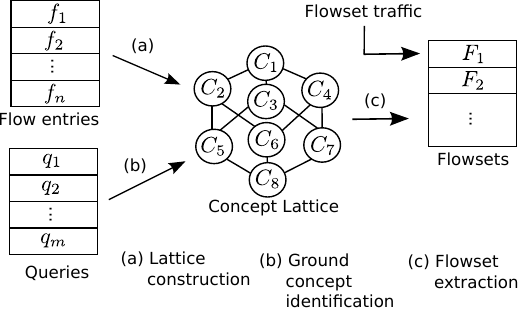}
	 \caption{Lattice based traffic measurement overview}
     \label{fig:FF}
\end{figure}

Following a novel statement of the traffic measurement problem, we introduce a set of easy-to-implement algorithms for building/maintaining lattices and counter structures and illustrate them with an example.
%
The global workflow is illustrated in Figure~\ref{fig:FF}. The lattice building algorithm (a)
outputs a structure that hierarchically organizes  flow entries. Flowset partition identification (b) and extraction (c) algorithms find optimal groups of flow entries based on user queries.


	\begin{table}[htb]
	\renewcommand{\arraystretch}{1.0}
	\caption{Notations}
	\label{tab:notations}
	\centering
		\begin{tabular}{ l l  | l l}
		\hline
		$\mathcal{F}$&	Set of supported flow entries &
		$\mathcal{H}$&	Set of matchfield values  \\
		$\mathcal{M}$&	Flow-to-matchfield incidence  &
		$\mathcal{K}$&	A Context $\mathcal{K}(\mathcal{F},\mathcal{H},\mathcal{M})$  \\
		$\mathcal{C_{\mathcal{K}}}$&	Context $\mathcal{K}$ concept set  &
		$I_c$&		Concept $c$ Intent  \\
		$E_c$&		Concept $c$ Extent  &
		$c{\widehat{~}}$&		Concept $c$ parent concepts  \\
		$v(c)$&	Concept c query vector   &
		$Q$&	Set of user queries\\
		$g(c)$&	Flows grounded in $c$&
		$\vert .\vert$& The cardinality of a set\\
		$t(c)$	&	Queries targeted at $c$  &
		$T$&	Set of Target concepts   \\
		$P$&	Set of Projection concepts  &
		$G$&	Set of Ground concepts  \\ \hline

		\end{tabular}
	\end{table}

\subsection{Definitions}
Below, a summary of the notions underlying our approach is provided (see~\cite{fca:ganter-99,fca-book:carpineto+04} for a complete coverage).
\begin{itemize}
\item A flow entry $f$ is defined by its set of matchfields $\lbrace h_1, h_2, .., h_n\rbrace$ and is assigned a counter to be updated whenever a packet matches $f$. Let $\mathcal{F}$ be the set of all flows installed in a specific switch while $F$ is flowset, i.e., an arbitrary set of flows (as in~\cite{Yuan:2011:PTP:1959441.1959451}). Let $\mathcal{H}$ be the set of all matchfield values from $\mathcal{F}$. 
\item A user query $q \in Q$ is a sequence of regular expressions on flow matchfields. Here, we assume a query is merely a set of matchfield values. 
\item A context $\mathcal{K}(\mathcal{F},\mathcal{H},\mathcal{M})$ associates $\mathcal{F}$ to $\mathcal{H}$ via an incidence relation  $\mathcal{M} \subseteq \mathcal{F} \times \mathcal{H}$. In ${\mathcal K}$ two \textit{image} operators $'$ lift $\mathcal{M}$ to the set level: flow/matchfield sets are mapped into the set of incident matchfields/flows (quantification is universal).
\item A concept is a pair $(F,H)$, where $F \in \wp(\mathcal{F})$ (\textit{extent}) and $H \in \wp(\mathcal{H})$ (\textit{intent}) are s.t. $F=H'$ and $H=F'$. The set $\mathcal{C}_\mathcal{K}$ of all concepts in ${\mathcal K}$ is partially ordered by extent inclusion:
$$ (F_1, H_1) \leq_{\mathcal K} (F_2, H_2) \Leftrightarrow  F_1 \subseteq F_2, (H_2 \subseteq H_1). $$
$\langle\mathcal{C}_\mathcal{K}, \leq_{\mathcal K}\rangle$ is a complete lattice, as meets $\wedge$ and joins $\vee$ are defined for arbitrary concept sets. The precedence $\prec_{\mathcal K}$, transitive reduction of $\leq_{\mathcal K}$, induces the Hasse diagram of the lattice. 
\item Compositions $''$ of complementary images $'$ are closure operators on $\wp({\mathcal F})$ and $\wp({\mathcal H})$, respectively. The families of extents, ${\mathcal C}^f_{{\mathcal K}}$, and of intents, ${\mathcal C}^h_{\mathcal K}$, are closed by $\cap$.
Thus, for a set $A$ of flows (of matchfields) $A''$ is the smallest extent (intent) comprising $A$.
\item $T$ is the set of target concepts: for a query $q$ its target is $\gamma(q)=(q',q'')$; $P$ is the set of projections, i.e., the meets of non-empty set of targets: $c_p = \bigwedge T_p$, $T_p \subseteq T$; $G$ is the set of ground projections: for a flow $f$ its ground is $\mu(f) = \min(\{(F,H) \in P | f \in F\})$.
\end{itemize}

\subsection{Problem statement}

The traffic measurement optimization problem consists in, given a set of flows $\mathcal{F}$ to monitor and a set of users queries $Q$, finding the minimal partition of $\mathcal{F}$ while being able to answer all user queries.
In this settings, the number of partitions in the classical approaches equals the number of flows, therefore, traffic measurement resources are maximal in all usage contexts.
Our running example, $\mathcal{K}(\mathcal{F},\mathcal{H},\mathcal{M})$, is shown in Table~\ref{tab:input-context}.
	\begin{table}[htb]
	\renewcommand{\arraystretch}{1.0}
	\caption{Input context}
	\label{tab:input-context}
	\centering
		\begin{tabular}{|c|c|c|c|c|c|c|c|c|c|c|}
		\hline
			&$h_1$	&$h_2$	&$h_3$	&$h_4$	&$h_5$	&$h_6$	&$h_7$	&$h_8$	&$h_9$	&$h_{10}$  \\ \hline
			$f_0$ &x   &   &   &x   &   &   &x   &   &x   &    \\ \hline
			$f_1$ &x   &   &   &x   &x   &   &x   &   &   &x    \\ \hline
			$f_2$ &   &x   &   &   &   &x   &   &x   &   &    \\ \hline
			$f_3$ &   &x   &   &   &   &x   &   &x   &   &x    \\ \hline
			$f_4$ &x   &   &   &   &   &   &x   &   &x   &    \\ \hline
			$f_5$ &x   &   &   &   &x   &   &x   &   &   &x    \\ \hline
			$f_6$ &   &   &x   &   &   &x   &   &   &x   &x    \\ \hline
			$f_7$ &   &   &x   &   &   &x   &   &x   &   &x    \\ \hline
		\end{tabular}
		\begin{tabular}{l l  l  l }
~\\
$h_1$ & - Ingress Port = 1 & $h_6$ & - IPv4 src = 132.208.130/32 \\
$h_2$ & - Ingress Port = 2 & $h_7$ & - IPv4 src =10/8 \\
$h_3$ & - Ingress Port = 3 & $h_8$ & - IPv4 dst = 10/8 \\
$h_4$ & - MAC src = $\mbox{MAC}_{1}$ & $h_9$ & - IPv4 dst = 132.208.130.1 \\
$h_5$ & - MAC dst = $\mbox{MAC}_{12}$ & $h_{10}$ & - Layer 4 dst port = 21 \\	
\end{tabular}
	\end{table}
%
\begin{figure}[htb]
     \centering
     \includegraphics[width=1.0\linewidth]{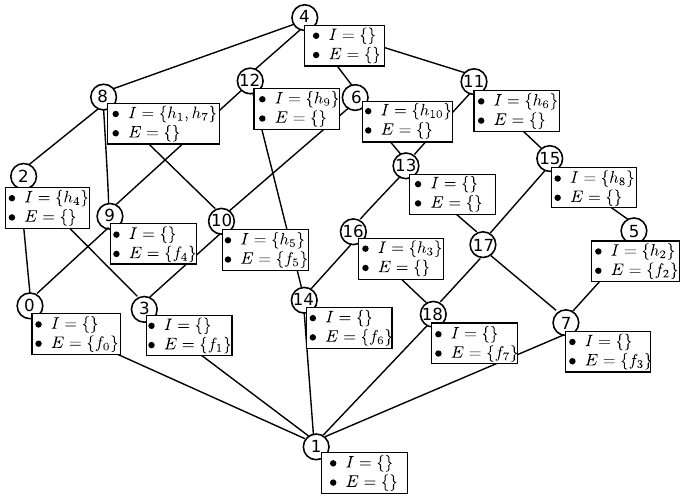}
	 \caption{Concept lattice of the input context: reduced concept intents/extents are provided to increase readability (objects are inherited upwards and attributes downwards)}
     \label{fig:CL}
\end{figure}

\subsection{Lattice construction}

%
\incmargin{1em}
\begin{algorithm}[!htb]

  \SetKwData{Left}{left}\SetKwData{This}{this}\SetKwData{Up}{up}
  \SetKwFunction{Union}{Union}\SetKwFunction{FindCompress}{FindCompress}
  \SetKwInOut{Input}{input}\SetKwInOut{Output}{output}
  \SetKwFunction{Sort}{Sort}

  \Input{$\mathcal{K}(\mathcal{F},\mathcal{H},\mathcal{M})$}
  \Output{Set of linked concepts $C$ }  
  
  $ConceptQueue \leftarrow \lbrace (\mathcal{F},\mathcal{F''})\rbrace$\;

%
  
  \While{$ConceptQueue \neq \varnothing$}{
  	$c = (F_c, H_c) \leftarrow ConceptQueue.pop()$\;
  	$Children \leftarrow \varnothing$\;
  	\ForEach{$h$ \textit{in} $\mathcal{H} - H_c$}{
  		$F_h = F_c \cap h'$\;
  		\If{$\exists \bar{c} \in  Children$ s.t. $ E_{\bar{c}} = F_h$}{
  			
			$E_{\bar{c}} \leftarrow E_{\bar{c}} \cup \lbrace h\rbrace$ \;
  		}\Else{
  			$Children \leftarrow Children \cup \lbrace(F_h,H_c \cup \lbrace h\rbrace)\rbrace$
  		}
  	}
  	$c{\widecheck{~}} \leftarrow \max(Children)$\;
  	$C \leftarrow C \cup c{\widecheck{~}}$\;
  }
  
  \caption{Lattice construction algorithm}
  \label{alg:construction}
\end{algorithm}\decmargin{1em}
Algorithm~\ref{alg:construction} is a version of \textit{NextNeighbor} in~\cite{fca-book:carpineto+04}, (p.35). It constructs the lattice from the top concept $({\mathcal F}, {\mathcal F}')$ down to the bottom one $({\mathcal H}', {\mathcal H})$, by generating the children of the current concept $(F,H)$. To that end, it first produces the extents of a larger set of sub-concepts by intersection of $F$ with the images of all matchfields outside of $H$. It then connects as children of $(F,H)$ only the maximums of the resulting set (and enqueues these for further processing). For instance, at concept $c_8 = (\{f_0, f_1, f_4, f_5\},\{h_1, h_7\})$ in Figure~\ref{fig:CL}, the following four extents are generated: $\emptyset$ (by intersections with $h_2'$, $h_3'$, $h_6'$, and $h_8'$), $\{f_0, f_1\}$ (with $h_4'$), $\{f_1, f_5\}$ (with $h_5'$ and $h_{10}'$), and $\{f_0, f_4\}$ (with $h_9'$). The latter three are maximal, hence they are the extents of children concepts for $c_8$ ($c_2$, $c_{10}$, and $c_9$, respectively). 

\subsection{Flowset partition identification}


\incmargin{1em}
\begin{algorithm}[!htb]
  \SetKwData{Left}{left}\SetKwData{This}{this}\SetKwData{Up}{up}
  \SetKwFunction{Union}{Union}\SetKwFunction{FindCompress}{FindCompress}
  \SetKwInOut{Input}{input}\SetKwInOut{Output}{output}

  \Input{A list of concepts $C$, \\
  		 A set of queries $Q=(q_1, .., q_i, .., q_n)$}
\Output{Target, projection and ground sets $(T, P, G)$ }  
  $Sort(C)$\;  
  
  \ForEach{$c$ \textit{in} $C$}{
  	\For{$q_i$ \textit{in} $Q$}{
	  	\If{$q_i \subseteq I_c$}{ 
	  		$v(c)[i] \leftarrow 1$\;
	  		$T \leftarrow T \cup \{c\}$;
	  		$Q \leftarrow Q - \{q_i\}$\;	  		
			}
		}	
	
	
	$v(c) \leftarrow v(c)\cup\bigcup_{\overline{c} \in c\widehat{~}} v(\overline{c})$\;
	\If{$\vert v(c)\vert>\mathsf{max}_{~\overline{c}~\in~c\widehat{~}~}(\vert v(\overline{c})\vert)$}{
		$P \leftarrow P \cup \{c\}$\; 
	}	
	\If{$\vert I_c \vert$=1}{
		\For{$p \in P$}{
			\If{$v(p)=v(c)$}{
			$G \leftarrow G \cup \{p\}$;
			$break()$\;
			
			}
		}		
	}
	}
	
  \caption{Flowset partition identification algorithm}
  \label{alg:ext}
\end{algorithm}\decmargin{1em}

The ultimate goal is to split  $\mathcal{F}$ into disjoint sets to be assigned a single counter each. Assume a query set $Q = \lbrace q_i\rbrace_{i=1..5}$ with $q_1=\lbrace h_{10}\rbrace$, $q_2 = \lbrace h_2, h_6, h_8\rbrace$, $q_3 = \lbrace h_1\rbrace$, $q_4 = \lbrace h_1, h_4, h_7\rbrace$  and $q_5 = \lbrace h_7\rbrace$.
Given the concept set $\mathcal{C_\mathcal{K}}$ and $Q$, Algorithm~\ref{alg:ext} parses $\mathcal{C_\mathcal{K}}$ to identify $T$, $P$ and $G$.	
Projection computation is supported by a bitvector whose value for $c=(F,H)$ reflects the queries satisfied by flows in $F$. Formally, the query vector $v(c)$ is an N-bit string indicating which $q_i$ are matched by $H$:
	$$
v((F,H))[i] = \left\{ \begin{array}{rl}
 1, &\mbox{ if $q_i\subseteq H$} \\
  0, &\mbox{ otherwise}
       \end{array} \right.
       1\leq i\leq N
$$
	\begin{table}[htb]
	\renewcommand{\arraystretch}{1.0}
	\caption{Query vectors values}
	\label{tab:v}
	\centering
\begin{tabular}{l l | l l}
\hline 
Query vector &  Concepts & Query vector &  Concepts\\ 
\hline 
$00000$  & $\mathbf{c_4}, c_{11}, c_{12}, c_{15}$ &
$00111$ &  $c_0,\mathbf{c_2}$ \\
$10000$  & $\mathbf{c_6}, c_{13},c_{14}, c_{16}, c_{17}, c_{18}$&
$10101$  & $\mathbf{c_{10}}$\\
$01000$  & $\mathbf{c_5}$&
$11000$  & $\mathbf{c_7}$\\
$10111$  & $\mathbf{c_3}$&
$00101$  & $\mathbf{c_8}, c_9$\\
$11111$ &  $\mathbf{c_1}$\\

\hline 
\end{tabular}
	\end{table}
	
First, the concepts list $C$ is sorted in decreasing order of extent sizes (line 1), to ensure the first concept whose intent matches a $q \in Q$ is its target (line 4). Matched $q$ are removed from the list (line 6). In our example, the algorithm outputs the targets $c_6, c_5, c_8, c_2$ and $c_8$, for $q_i$ ($i=1..5$), respectively.
%
Then, the value of $v(c)$ is finalized (line 7): the local part (targeted queries, line 5) is merged with the inherited parent values (see results in Table~\ref{tab:v}). Projections are concepts whose query vectors have more 1s than any of their respective parent ones (line 8). For instance, $c_{10}$ has three 1s, more than its parents $c_8$ (one) and $c_6$ (two), hence it is a projection (as meet of the targets $c_6$ and $c_8$).
Finally, the ground concept of a $f \in \mathcal{F}$ is the projection with the same query vector as the flow concept $(f'',f)$ (lines 10-13). Table~\ref{tab:ground} provides the flow-to-ground mapping of our example.

	\begin{table}[!t]
	\renewcommand{\arraystretch}{1.0}
	\caption{Flow-to-ground concept mapping}
	\label{tab:ground} 
	\centering
	\begin{tabular}{l l | l l | l l | l l}
	\hline 
	Flow &  Ground & Flow &  Ground & Flow &  Ground & Flow &  Ground  \\ \hline
	$f_0$ & $c_2$ &
	$f_1$ & $c_3$ &
	$f_2$ & $c_5$ &
	$f_3$ & $c_7$ \\
	$f_4$ & $c_8$ &
	$f_5$ & $c_{10}$ &
	$f_6$ & $c_6$ &
	$f_7$ & $c_6$ \\
	
	\hline 
	\end{tabular}
	\end{table}

%
%

%

\subsection{Flowset extraction}


The optimal partition is composed by the target concept extents:
$\Phi_g = \lbrace F \vert \exists (F,H) \in G \rbrace$.
%
A hardware counter is assigned to each flowset in $\Phi_g$, i.e., a total of $m = \vert \Phi_g \vert$ counters. And since ground intents are disjoint, $m \leq \vert\mathcal{F}\vert$ with $=$ reached with exclusively singleton flowsets. With $G$ from Table~\ref{tab:ground}, $f_6$ and $f_7$ share a common counter, whereas the remaining flows get a dedicated counter each. 

\subsection{New flow entry insertion}
\incmargin{1em}
\begin{algorithm}[!htb]
  \SetKwData{Left}{left}\SetKwData{This}{this}\SetKwData{Up}{up}
  \SetKwFunction{Union}{Union}\SetKwFunction{FindCompress}{FindCompress}
  \SetKwInOut{Input}{input}\SetKwInOut{InputOutput}{input/output}
  \SetKwFunction{Sort}{Sort}
  \Input{Added flow entry $f_n$}
  \InputOutput{Concept lists $(C,T,P,G)$}  
  	
    $C_n \leftarrow \varnothing$; 
  $M \leftarrow \varnothing$\; 
	\ForEach{$c = (E_c,I_c)$ \textit{in} $C$}{
	$H \leftarrow I_c \cap f_n'$\;
	     	\If{$H = I_c$}{ 
     		$c \leftarrow (E_c \cup \lbrace f_n\rbrace,I_c)$;
     		$M \leftarrow M \cup \{c\}$\;
     	}\ElseIf{$\nexists \bar{c} \in C_n \cup M$ s.t. $I_{\bar{c}} = H$}{ 
     		$C_n \leftarrow C_n \cup \{c_n = (E_c \cup \lbrace f_n\rbrace,H)\}$\;
     		$c\widehat{~} \leftarrow c\widehat{~} \cup \{c_n\}$; $c_n\widecheck{~} \leftarrow c_n\widecheck{~} \cup \{c\}$\;
     		$UpdateOrder(c_n,c,C_n,M)$\;
     		UpdateStatus$(c_n, c, T, P, G)$\;
     	
%
%

%
	}
	}
	$c_p \leftarrow Lookup(P, v(\mu(f_n))$;\CommentSty{  //Ground of $f_n$} \\
	$g(c_p) \leftarrow g(c_p) \cup \{f_n\}$\;
	$G \leftarrow G \cup \{c_p\}$\;
	$C \leftarrow C \cup C_n$\;
  \caption{Lattice update : Add a flow}
  \label{alg:add-flow}
\end{algorithm}\decmargin{1em}

\begin{figure}[ht]
	     \centering
	     \includegraphics[width=1.0\linewidth]{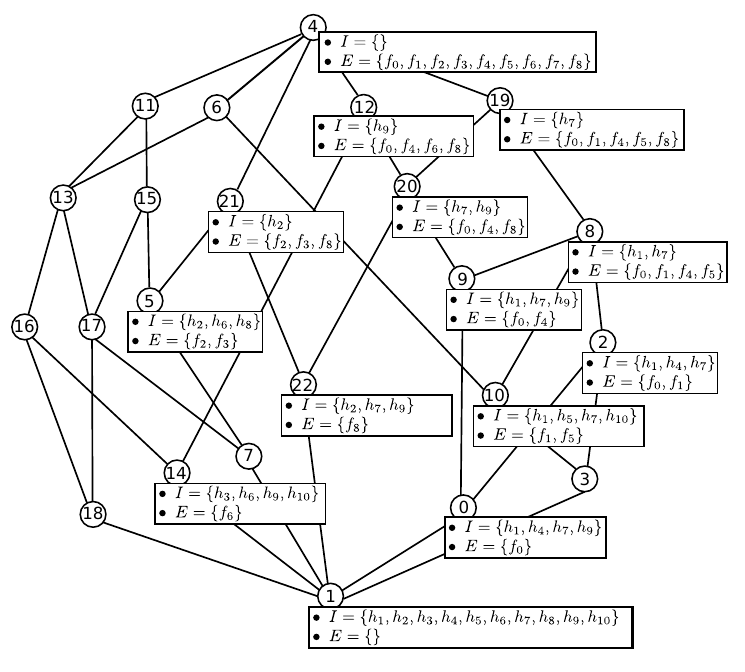}
		 \caption{The concept lattice of the extended context: only full intents/extents of relevant concepts are drawn.}
	     \label{fig:FCL-inc1}
	 \end{figure}

Assume a new flow $f_8$ with $f_8'=\{h_2, h_7, h_9\}$ is added to the initial context. Algorithm~\ref{alg:add-flow} implements the schema in~\cite{fca-algo:valtchev+03} to update the lattice ${\mathcal L}_{\mathcal K}$ to the lattice of ${\mathcal K}_n = (\mathcal{F}_n,\mathcal{H},\mathcal{M}_n)$, where $ \mathcal{F}_n = \mathcal{F} \cup \{f_n\}$ and $ \mathcal{M}_n = \mathcal{M} \cup \{f_n\} \times f_n'$. The basic task consists of producing all intersections of the new flow image $f_n'$ with intents from ${\mathcal C}_{\mathcal K}$. For intersections that are intents in ${\mathcal K}$, the extent of the underlying concept (qualified as \textit{modified}) is updated with $f_n$ (line 5). For instance, $c_{12}$ yields	$I_{c_{12}} \cap f_8 ' = I_{c_{12}} $, thus its extent is updated with $f_8$ (see the updated lattice in Figure~\ref{fig:FCL-inc1}). The only other modified is $c_4$. Figure~\ref{fig:FCL-inc1} presents the updated lattice in Figure~\ref{fig:CL}. Observe that concept numbers are IDs: concepts with the \textit{same numbers} as in Figure~\ref{fig:CL} have the \textit{same intents}.


	 	
An intersection missing in ${\mathcal C}^h_{\mathcal K}$ triggers the creation of a new concept (only the first time). The intent of $c_n$ is the intersection itself, while the extent is the extent of the generating concept (alias the \textit{genitor}) plus $f_n$ (line 7). In our example,  $c_{19}$, $c_{20}$, $c_{21}$, and $c_{22}$ are the new concepts with genitors $c_8$, $c_9$, $c_5$, and $c_1$, respectively. Among them, $c_{22}$ is the flow concept of $f_8$.

%
%
	
To update the $\prec_{\mathcal{K}}$ links, first $c_n$ and its genitor are linked as a parent and a child, respectively (line 8). Then, the children of $c_n$ are chosen among the already identified part of the new and modified concept sets (line 9).

	 
%
	

\incmargin{1em}
\begin{algorithm}[!htb]
  \SetKwData{Left}{left}\SetKwData{This}{this}\SetKwData{Up}{up}
  \SetKwFunction{Union}{Union}\SetKwFunction{FindCompress}{FindCompress}
  \SetKwInOut{Input}{input}\SetKwInOut{InputOutput}{input/output}
  \SetKwFunction{Sort}{Sort}
  \InputOutput{concepts $c$, $c_n$ (the new and its genitor) \\Concept sets $(T,P,G)$}

	\ForEach{$q_i \in t(c)$}{
		\If{$q_i \subseteq H$}{
			$t(c_n) \leftarrow t(c_n) \cup \{q_i\}$;
			$t(c)\leftarrow t(c) - \{q_i\}$\;
			$v(c_n)[i] \leftarrow 1$\;
		}
	}
	\lIf{$t(c_n) \neq \varnothing$}{$T \leftarrow T \cup \{c_n\}$}\;
	\lIf{$t(c) = \varnothing$}{$T \leftarrow T - \{c\}$}\;
	\If{$\vert v(c_n)\vert>\mathsf{max}_{~\overline{c}~\in~c_n^{\widehat{~}}~}(\vert v(\overline{c})\vert)$}{
		$P \leftarrow P \cup \{c_n\}$\;
		\If{$\vert v(c)\vert = \vert v(c_n)\vert$}{
			$P\leftarrow P - \{c\}$\;	
			
			\If{$g(c_n)\neq \varnothing$}{
			$g(c_n) \leftarrow g(c)$;
			$g(c) \leftarrow \varnothing$\;
		
			$G\leftarrow G \cup \{c_n\} - \{c\}$\;
			}
		}
	}

%

  \caption{UpdateStatus: Update support structure}
  \label{alg:ctr-update}
\end{algorithm}\decmargin{1em}
The UpdateStatus method, detailed by Algorithm \ref{alg:ctr-update}, establishes the status of the new concept $c_n$ (target, projection, ground, none) and updates that of its genitor $c$. In our example, $c_{19}$ matches $q_5=\{h_7\}$, thus, $q_5$ will be re-targeted from the genitor $c_8$ to the new concept $c_{19}$ (line 3). The projection test (line 7) follows the one in Algorithm~\ref{alg:ext}. Next, if the query vector of $c_n$ has the same number of 1s as the vector of $c$ (line 7), the latter is no more a projection (line 8). In this case, all flows grounded at $c$ are re-grounded at $c_n$ (lines 11-13).

Finally, in a post-processing step (lines 11-13, Algorithm~\ref{alg:add-flow}) the ground concept of the new flow $f_n$ is established as the projection $c_p \in P$ with the same query vector as the flow concept $\mu(f_n)$ (line 11). 	
In our example, $v(\mu(f_8)) = v(c_{22}) = 00001$ is the same as the projection $c_{19}$ query vector. The new flow $f_8$ is grounded to $c_{19}$ (line 13).

To sum up the restructuring: The new target concept list is $T_n = T \cup \lbrace c_{19}\rbrace$, the new projections $P_n = P \cup \lbrace c_{19}\rbrace$, and the new grounds $G_n = G \cup \lbrace c_{19}\rbrace$.

\vspace{.2cm} 
Our solution comprises algorithms for removing of flows as well as for adding/removing queries or individual matchfield values that we do not provide here for space limitation reasons.
All our algorithms follow a similar computational schema: they perform a traversal of the concept lattice whereas the most effort-intensive processing task for a concept boils down to one set-based operation on the intents/extents of each parent/child. Hence the algorithmic complexity is $O(|{\mathcal C}_{\mathcal K}| * |\mathcal{F}|* (|\mathcal{H}| + |\mathcal{F}|))$~\cite{fca-algo:valtchev+03}. Below, we show how the richness and regularity of the lattice and its counter-related substructure translate into algorithmic efficiency and optimality of the solution.

\section{Proofs}
\label{sec:proofs}

We establish below the correctness of our algorithms (Properties~\ref{prop:def-inc-comp} to~\ref{prop:proj-new-no-old}
and Theorem~\ref{thm:assign-corr}) and the minimalness of the proposed counter assignment (Theorem~\ref{thm:assign-min}).

\subsection{Lattice construction and maintenance}

Algorithm~\ref{alg:construction} 
is rooted in the following results~\cite{dm-fca:valtchev+04}: (1) For a concept $(F,H)$ with children $(F_i, H_i)$, the \textit{faces} $H_i - H$ are pairwise disjoint, and (2) $\forall h \in H_i - H$, $h' \cap F = F_i$.
Hence it constructs all $(F_i, H_i)$ from $(F,H)$ by producing all possible $h' \cap F$ for matchfields $h \in {\cal H} - H$ and partitioning those h into the faces of $(F,H)$. Non face matchfields from ${\cal H} - H$ generate smaller intersections that are filtered out.


Algorithm~\ref{alg:add-flow} (lattice update) transforms ${\mathcal C}_{\cal K}$ into ${\mathcal C}_{{\mathcal K}_n}$ with the emphasis on the completion of the intent family ${\mathcal C}^h_{\mathcal K}$ to ${\mathcal C}^h_{{\mathcal K}_n}$ (since ${\cal H}_n = {\cal H}$). 
Indeed, existing intents provably remain valid in ${\mathcal K}_n$: ${\mathcal C}^h_{\mathcal K} \subseteq {\mathcal C}^h_{{\mathcal K}_n}$. Moreover, the new intents are pairwise intersections of $f_n'$ with elements from ${\mathcal C}^h_{\mathcal K}$:

%
%
%
\begin{property}
\label{prop:def-inc-comp}
${\mathcal C}^h_{{\mathcal K}_n} = {\mathcal C}^h_{{\mathcal K}} \cup \{ \{f_n\}' \cap H \mid H \in {\mathcal C}^h_{{\mathcal K}} \}$.
\end{property}
Correspondingly, the extents of concepts in ${\mathcal C}_{{\mathcal K}_n}$ have only two possible forms: $F$ or $F \cup \{f_a\}$ where $F \in {\mathcal C}^f_{\mathcal K}$.

A new intent generates a new concept: Although it may be produced more than once, a canonical generator, the \textit{genitor}, exists that holds a two-fold bound to the new concept, i.e., both through its intent and extent:
\begin{property}
\label{prop:new-gen}
For a $(F,H) \in {\mathcal C}_{{\mathcal K}_n}$, s.t. $H \in {\mathcal C}^h_{{\mathcal K}_n} - {\mathcal C}^h_{\mathcal K}$, $\exists (F_g,H_g) \in {\mathcal C}_{\cal K}$ s.t.
$H = H_g \cap f_n'$ and $F = F_g \cup \{f_n\}$.
\end{property}
As a corollary, the genitor intent is the closure of the new one in ${\mathcal K}$: $H'' = H_g$.

Modified concept intents $H$ are provably s.t. $H \subseteq \{f_n\}'$ (closed in ${\mathcal K}$). Hence in ${\mathcal K}_n$ the respective extents $H'$ comprise $f_n$. Observe that genitor and modified have intents that are the closures of their $H \subseteq \{f_n\}'$, hence they are the maximal concepts to produce it. Thus, they will be the first ones to reach along the top-down breadth-first traversal of the lattice.
Finally, the adjustment of the new precedence among concepts in ${\mathcal C}_{{\mathcal K}_n}$ is skipped here  (interested readers are directed to~\cite{dm-fca:valtchev+04}).

\subsection{Measurement support construction}
Observe $\gamma(q)$ is the maximal $(F,H) \in {\mathcal C}_{\mathcal K}$ s.t. $q \subseteq H$ while $F$ is the set of all flows $f$ satisfying $q$ ($q \subseteq f'$). 
Moreover, $\mu(f)$ is well defined: it is the meet of the targets of queries satisfied by $f$ ($ \mu(f) = \bigwedge \{ \gamma(q) | q \in Q;~q \subseteq f' \}$). 

The main tasks in Algorithm~\ref{alg:ext} are detecting all $\gamma(q)$ (the highest concept $(F,H)$ with $q \subseteq H$) and propagating the targeted $q$ downwards in the lattice. These $q$ are stored in the bitvectors $v()$ for further projection tests.
Now, a projection $c$ is exactly the meet of the targets of queries in $v(c)$: 
\begin{property}
\label{prop:role-vect}
$c\in P$ iff $c = \bigwedge \{\gamma(q_i)~|~v(c)[i]=1 \}$. 
\end{property}
As a corollary, the projection extent is the intersection of the target ones ($F = \bigcap \{E_{\gamma(q_i)} | v(c)[i]=1 \}$). Then, $c$ is maximal for  $v(c)$ and thus can be recognized by comparing its bitvector to those of parent concepts:
\begin{property}
\label{prop:charact-proj}
$c \in P$ iff $\forall \bar{c} \in c{\widehat{~}}, v(\bar{c}) \neq v(c)$. 
\end{property}
As it is readily shown that as a function $v()$ is monotonously non increasing w.r.t. $\leq$, the property may be recast in terms of cardinalities: $|v(c)| > \mathsf{max}_{\bar{c} \in c{\widehat{~}}}(|v(\bar{c})|)$.

Finally, $G$ is tested by comparing $v(c)$ to bitvectors of flow concepts:
\begin{property}
\label{prop:charact-ground}
For a $c=(F,H)$, $c \in G$ iff $\exists f \in F$ s.t. for $\bar{c}=(f'',f')$, $v(\bar{c}) = v(c)$. 
\end{property}
Moreover, as in our specific case, no flow has a subset of another flow's matchfields, $\forall f \in {\mathcal F}$, $f''=f$. Thus flow concepts are exactly those with singleton extents.

\subsection{Measurement support maintenance}

To show that $T$, $P$ and $G$ are correctly transformed into $T_n$, $P_n$ and $C_n$, respectively, by Algorithm~\ref{alg:ctr-update}, observe that for $c \in {\mathcal C}_{\mathcal K}$ $v(c)$ keeps its value in ${\mathcal C}_{{\mathcal K}_n}$. 
\begin{property}
\label{prop:vect-preserve}
For a concepts $c=(F,H) \in {\mathcal C}_{{\mathcal K}_n}$, if $H \in {\mathcal C}^h_{\mathcal K}$ then $v_n(c) = v(\bar{c})$ where $\bar{c} = (H',H)$. 
\end{property}
The reason is $v(c)$ only depends on $H$ and $Q$ which remain stable in ${\mathcal K}_n$. Thus, the function $v_n()$ evolves from $v()$ by merely computing the values for new concepts in $C_n$  (value propagation matches the downward generation of $C_n$). 

Now, $T_n$ may depart from $T$ as some $q \in Q$ may change targets ($\gamma(q) \neq \gamma_n(q)$). Clearly, $\gamma_n(q)$ can only be a  new concept in ${\mathcal K}_n$, whereas $\gamma(q)$ is its genitor in ${\mathcal K}$.
%
%
\begin{property}
\label{prop:target-inc}
For a query $q \in Q$ s.t. $\gamma(q) \neq \gamma_n(q)$, $\bar{c}=\gamma_n(q)$ is a new concept in ${\mathcal C}_{{\mathcal K}_n}$ while $c=\gamma(q)$ is its genitor. 
\end{property}

This follows from the minimalness of $H''$ among intents comprising $H$. Thus, with $c=(F,H)$ and $c=(F_n,H_n)$, we show that $H=H_n''$ (hence the genitor status) in ${\mathcal K}$. Indeed, assuming $H \neq H_n''$, we deduce $H_n'' \subset H$ (*) since $q \subseteq H$ (recall $q'' = H$) and $H_n$ is the closure of $q$ in ${\mathcal K}_n$ (minimal intent comprising $q$). Yet since $q \subseteq H_n \subset H_n''$, (*) would contradict the minimalness of $H = q''$ in ${\mathcal K}$.

$P_n$ evolves from $P$ along two separate scenarios: (1) as with $T_n$, a new concept may become projection by eclipsing its genitor in $P_n$; and (2) a new concept may become the infimum for a set of query targets with no equivalent in $P$. Recall that projections are identified within $P$ by $v()$ (Property~\ref{prop:role-vect}). In other terms, in case one, the infimum $c$ of a set of targets ($v(c)$) evolves to a different concept $\dot{c}$ (diverging intents) with the same bitvector value ($v(c)=v_n(\dot{c})$), whereas in case two, a previously nonexistent set of targets $v_n(\dot{c})$ arises.   

\begin{property}
\label{prop:proj-new-old}
Given a $c\in P_n$, if $c$ is not the equivalent of the projection concept $\bar{c}= \bigwedge \{ \gamma(q_i)~|~v_n(c)[i]=1 \}$ in ${\mathcal K}$, then $c$ is a new concept with $\bar{c}$ as its genitor. 
\end{property}

Assume that for some $c=(F,H) \in P_n$, the projection $\bar{c}= \bigwedge \{ \gamma(q_i) | v_n(c)[i]=1 \}$ from $P$ is such that $\bar{c}=(F_o, H_o)$ and $H_o \neq H$ ($\bar{c}$ not an equivalent concept in ${\mathcal K}$, despite $v(c) = v(\bar{c})$). The latter means $F \neq F_o$, and since these are the intersection of the target extents from $v(c)$, it follows that \textit{all those extents} have changed. As we saw previously, the only possible evolution of a target extent for a query $q$ in ${\mathcal K}_n$ is to increase by $f_n$.
Consequently, their intersection $F$ (corollary of Property~\ref{prop:role-vect}) comprises $f_n$ as well and, as $H$ is not an intent in ${\mathcal K}$, $c$ is a new concept. By the same argument, $F_o$ can only be $F_o = F - \{f_n\}$, hence $\bar{c}$ is the genitor of $c$ (Property~\ref{prop:new-gen}).

In case two, the new projection $c$ is a new concept too:
\begin{property}
\label{prop:proj-new-no-old}
Given a $c=(F,H) \in P_n$, if for all projections $\bar{c} \in P$, $v_n(c) \neq v(\bar{c})$, then $c$ is a new concept. 
\end{property}

Assuming the opposite, let $H \in {\mathcal C}^h_{\mathcal K}$, hence $c$ is not a new concept ($f_n \notin F$) and thus $v_n(c) = v(c)$. Consequently there is a concept with the same bitvector value in ${\mathcal K}$, $c$ itself, which further means there must be a maximal concept $\bar{c}$, s.t. $v(c) = v(\bar{c})$, i.e., an infimum. This contradicts the starting hypothesis.

$G_n$ being a subset of $P_n$, similar evolution patterns hold: In the above case one, all flows grounded at the genitor --which vanishes from $P_n$, hence from $G_n$-- must be re-grounded at the new projection $c_n$. In case two, no flow from ${\mathcal F}$ could be grounded in $c_n$, since their flow concepts in ${\mathcal C}_{{\mathcal K}_n}$ have intents from ${\mathcal C}^h_{\mathcal K}$. Thus, the respective bitvectors do not change in ${\mathcal K}_n$, hence all such flows are grounded in $c$ whose $v_n(c)$ existed in $P$. Therefore, $f_n$ is the only candidate for case two grounds.


To sum up, in $T_n$, new concepts of target genitors grab targeted queries comprised in their respective intents, while genitors with no remaining queries vanish. In $P_n$, new concepts are tested for projection and, if positive, genitors too. In $G_n$, flows grounded at a shifting projection move from the genitor to the new concept. Finally, $\mu(f_n)$ is found. 

%

\subsection{Correctness and minimalness of counter assignment}

We prove that ground concept-based counter assignment is: (1) correct, and (2) of minimal cardinality. Recall that each ground $c_g \in G$ is assigned a counter whose support is the set of grounded flows denoted $g(c_g) = \{f | \mu(f) = c_g\}$. This is a unique counter assignment (uniqueness of $\mu(f)$) and w.l.o.g. we assume that each flow is grounded. 
Furthermore, for each $q \in Q$ the set of relevant counters compose to a sum and let the underlying total set of flows be $S(q)$. As a counter enters a query sum iff its ground is below the corresponding target, we have $ \forall f \in {\mathcal F}, q \in Q$,  $f \in S(q)$ iff $\mu(f) \leq \gamma(q)$.


Correctness means a $S(q)$ is the set of flows satisfying $q$:
\begin{theorem}
\label{thm:assign-corr}
$\forall q \in Q, f \in {\mathcal F}$, $q \subseteq f'$ iff $\mu(f) \leq \gamma(q)$.
\end{theorem}

'If': follows from intent inclusion along $\leq$: $q \subseteq I_{\gamma(q)} \subseteq I_{\mu(f)} \subseteq f'$.
'Only if': Observe that satisfaction means $f \in E_{\gamma(q)}$ and assume, by \textit{reductio ad absurdi}, $\mu(f) \not\leq \gamma(q)$.
Then the infimum $c_{q,f} = \mu(f) \wedge \gamma(q) \in P$ (as $\wedge$ is associative) whereas $f \in E_{c_{q,f}}$. Yet this contradicts $\mu(f) \not\leq \gamma(q)$
as then $c_{q,f} < \mu(f)$ (minimal in $P$ to hold $f$).

Conversely, redundancy in $S(q)$ is excluded since a relevant flow $f$ appears exactly once in it (through $\mu(f)$).

Minimalness means no unique counter assignment among a smaller set of counters could answer all $q$ in $Q$. We focus on the underlying partition of ${\mathcal F}$:
\begin{theorem}
\label{thm:assign-min}
Let $cpt : {\mathcal F} \rightarrow \wp({\mathcal F})$ with $cpt(f) = F$ iff $f \in F$ and assume $|\mathsf{ran}(cpt)| < |G|$. Then $\exists q \in Q$ s.t. $S(q)$ is not decomposable into the union of some sets from $\mathsf{ran}(cpt))$.
\end{theorem}

By \textit{reductio ad absurdi}, assume all $S(q)$ represent unions of $cpt(f)$ for flows $f$ from a well-chosen set.
A straightforward combinatorial argument yields $\exists f_1, f_2 \in {\mathcal F}$, s.t. $\mu(f_1) \neq \mu(f_2)$ (**) yet $cpt(f_1)  = cpt(f_2)$.
Yet (**) means $v(\mu(f_1)) \neq v(\mu(f_2))$ and w.l.o.g. we can assume $\exists q_a \in Q$, s.t. $q \subseteq f_1'$ but $q_a \not\subseteq f_2'$.
However, this contradicts the initial hypothesis since there is no way to correctly decompose $S(q_a)$ into a union of $cpt(f)$: if $cpt(f_1)$ participates, then there is no way to remove the contribution of $f_2$ (subtraction not available) while otherwise, there is no way to recover the contribution of $f_1$ ($cpt(f_1)$ is its unique counter).

\section{Implementation and Results}
\label{sec:results}



A FlowME implementation over an OpenFlow switch was studied along three measurement axes: (1) memory cost expressed in term of number of managed counters, (2) processing effort for concept lattice generation/update, and (3) performance overhead on packet processing. 
%
%
We choose OpenFlow because flow entries and user queries can be pushed and retrieved from the Openflow tables of the switch.
%
%
Experimental results show a huge reduction in the number of hardware counters with a reasonable overall computational effort and negligible interference on traffic.


\subsection{Testbed design}
\begin{figure}[ht]
	     \centering
	     \includegraphics[scale=2.2]{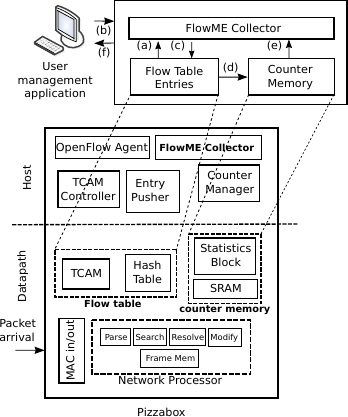}
		 \caption{FlowME testbed}
	     \label{fig:FTestBed}
	 \end{figure}

The FlowME testbed comprises an OpenFlow Switch with per-flow counter support, a flow entry generator, a Collector and user applications that generate queries. 
 As shown in Figure~\ref{fig:FTestBed}, FLowME Collector gets the set of flow entries $\mathcal{F}$ (a) installed in the Flow table of the OpenFlow switch  and user queries $Q$ (b). It calls upon lattice algorithms of the Coron FCA suite~\cite{coron-url} to calculate/maintain the optimal flow entry partition and exploits it to place flow counter references (c). Next, traffic matching $\mathcal{F}$ increments hardware counters (d). FlowME collector reads counter values (e), calculates query answers and sends them to user applications (f). We experimented FlowME with a variety of flow entry and query distributions. 

\subsubsection{Flow entry benchmarking}

Flow entry benchmark generates up to 12 fields per entries. It is based on Flexible Rule Generator~\cite{5682304}, a user controlled benchmarking tool for evaluating packet forwarding algorithms that generate sets of OpenFlow flow entries based on predefined matchfield distributions. 
We extract matchfield distributions from packet traces provided by packetlife.net. The traces are particularly interesting since packet headers contain different types of fields as MAC, VLAN, IP and transport fields. As a result, a total of 12 standard OpenFlow matchfields are used in the benchmark.
We analyse packet trace headers to determine a distribution function for each matchfield. 
Table~\ref{tab:mfvd} shows a set of matchfields and their value distributions. Notice that IP source and destination analysis involves the prefix distribution and the prefix length distribution.
	\begin{table}[htb]
	\renewcommand{\arraystretch}{1.0}
	\caption{Packet trace matchfield value distribution of density $\geq 3\%$}
	\label{tab:mfvd}
	\centering
		\begin{tabular}{|c|c|c|c|c|}
		\hline
		\textbf{Matchfield}	&	\textbf{Distribution}	\\ \hline
		MAC src				&	00:40:05(39\%), 08:00:07(13\%), 00:60:08(19\%) \\ \hline 
		MAC dst				&	00:60:08(33\%), FF:FF:FF(37\%), 00:40:05(19\%) \\ \hline
		Ethertype			&	0x8100(98)\% \\ \hline
		VLAN id				&	32(56\%), 104(17\%), 108(4\%), 6(6\%)\\ \hline
		IP protocol			&	0x06(80\%), 0x11(6\%), 0x01(13\%) \\ \hline
		TOS					&	0(96\%), 192(3\%) \\ \hline
		L4 src port			&	2212(41\%, 1815(26\%), 2388(11\%), 8(4\%) \\ \hline
		L4 dst port		&	1815(53\%), 2212(18\%), 2388(8\%), 3314(4\%) \\ \hline
		\end{tabular}
	\end{table}

\subsubsection{Query benchmarking}	

The second benchmark generates application queries. 
A query covers a set of flow entries whose size depends on how many matchfields get a non wildcard value.
In our experimental study, we generate user queries with the same matchfield value distribution as flow entries, in which we inserted some wildcarded values (a specific percentage for each matchfield).
Moreover, we force each query to cover at least one flow entry. To that end, we first extract $n$ flow entries from $\mathcal{F}$ and then insert a specific percentage of wildcards in each matchfield, thus yielding a set of $n$ queries.

\subsection{Switch implementation}
	 
\begin{figure}[ht]
	     \centering
	     \includegraphics[scale=2.5]{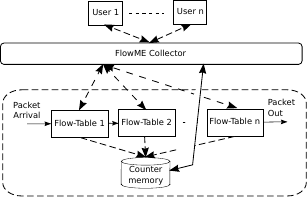}
		 \caption{Implementation over an OpenFlow switch
		 }
	     \label{fig:EZimp}
	 \end{figure}	 
We use the 100 Gig OpenFlow implementation over EZchip's network processor introduced in~\cite{PerformanceICC2012}, and we add flow counter support. In our solution, each flow in memory is split across a set of pipelined tables (see Figure~\ref{fig:EZimp}). Each table is implemented using TCAM for flow matchfield classification and Hash table for flow entry instruction and counter reference storage. 
Flow tables report counters in a continuous memory space where SRAM supports first 8192 counters and RLDRAM supports the rest.
In a typical scenario, as the one shown in Figure~\ref{fig:FTestBed} and~\ref{fig:EZimp}, when a packet is received, relevant header fields are parsed by a Parse engine, and a key is built. A lookup is performed by a Search engine in the TCAM against flow entry matchfields, and if they match, the TCAM provides an index of the matching flow entry. In order to retrieve counters and instructions associated to that entry, a second lookup is performed in Hash table based on the entry index. A Resolve engine receives the entry and processes its hardware counter reference and instructions (prepared for an eventual execution). The processing repeats on subsequent flow tables. At the very end of flow identification, Resolve engine sends a hardware counter increment command to the Statistics Block via a dedicated routine. 

\subsection{Memory cost}
\label{sec:mem-cost}
\begin{figure}[ht]
	     \centering
	     \includegraphics[scale=0.55]{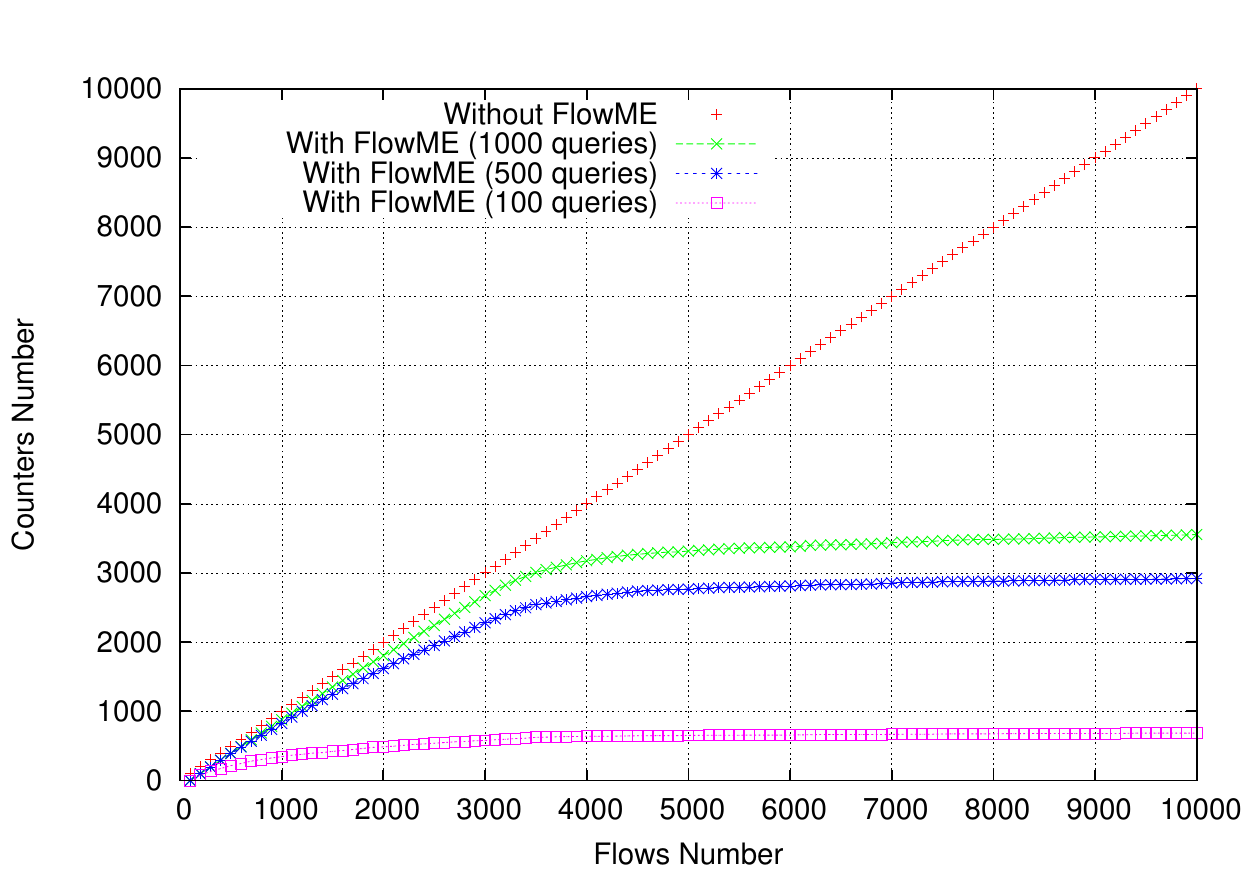}
		 \caption{Number of managed counters for varying query set sizes. Query field value distribution: 10\% exact-match 90\% wildcard}
	     \label{fig:FMS09}
	 \end{figure}

\begin{figure}[ht]
	     \centering
	     \includegraphics[scale=0.55]{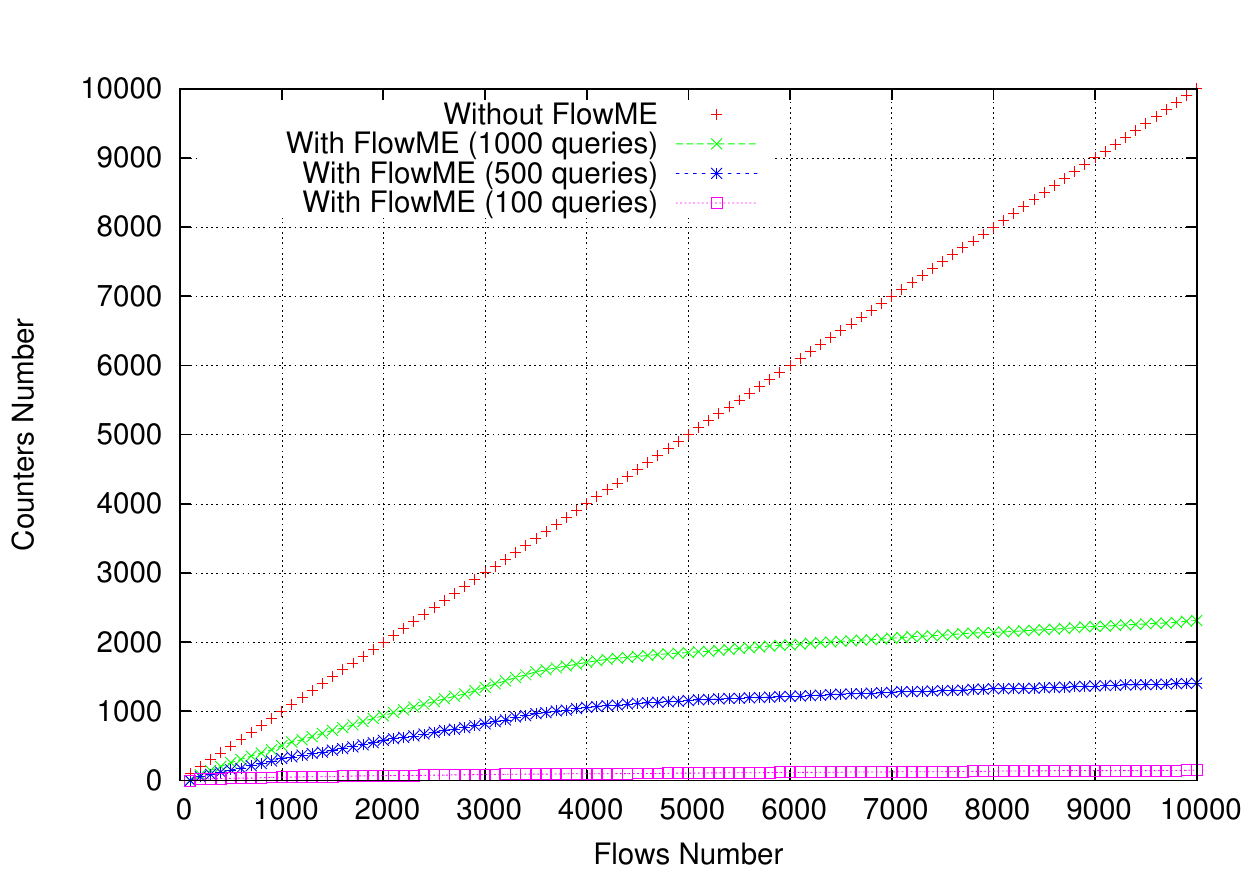}
		 \caption{Number of managed counters for varying query set sizes. Query field value distribution: 50\% exact-match 50\% wildcard}
	     \label{fig:FMS05}
	 \end{figure}

\begin{figure}[ht]
	     \centering
	     \includegraphics[scale=0.55]{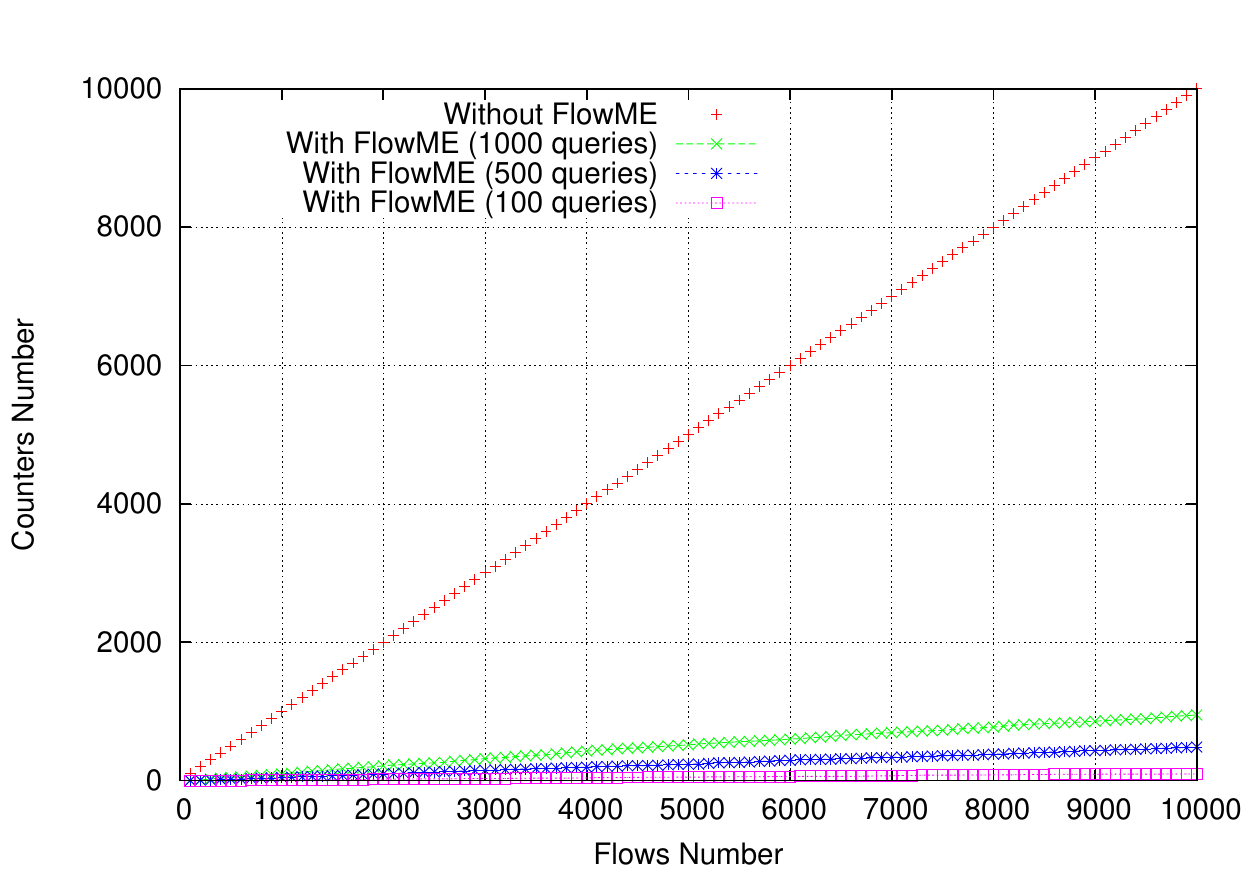}
		 \caption{Number of managed counters for varying query set sizes. Query field value distribution: 90\% exact-match 10\% wildcard}
	     \label{fig:FMS01}
	 \end{figure}

Per-flow traffic measurement tools manage an individual counter for each traffic flow processed by the system (the set $\mathcal{F}$) and report individual flow statistics to a centralized collector. In those systems, the counter number $N_c$ evolves linearly with $\vert\mathcal{F}\vert$ since, in practice, for each flow $f \in \mathcal{F}$, the system may need different traffic metrics, e.g. the number of matching packets or their total size. 
In contrast, our solution relies on aggregated counters. so we ran FlowME with the above benchmarks and observed the $N_c$ value.
The evolution of the number of counters to maintain in order to answer a set of user queries $Q$ of size $N_Q$ is depicted in Figures~\ref{fig:FMS09}, \ref{fig:FMS05} and~\ref{fig:FMS01}.
In Figure~\ref{fig:FMS09}, query field values are composed of 10\% exact match values and 90\% wild-cards. In Figures \ref{fig:FMS05} and~\ref{fig:FMS01}, queries are more specific with 50\% and 90\% exact match values, respectively. 
For example, the first experiment shows that for $N_Q=1000$, $N_c$ is significantly lower than the 10000 per-flow counters of the base-line solution (949 to 3555, depending on wild-card distribution). As a general trend, we observe that as less specific queries cover more flow entries each, the number of (minimal) intersections is higher and thus the counter set grows larger.
Let $M$ be the available memory size reserved for traffic measurement, $M_u$ the set of managed counter registers in a specific measurement period. $M_u$ is bounded by the number of  flow entries installed in the switch. In traditional flow-based measurement schemes, $M_u = \vert\mathcal{F}\vert$. In contrast, FlowME is an application-aware measurement tool, hence $\vert\mathcal{F}\vert$ is the worst case that occurs only if the ground intersections generated by $Q$ split $\mathcal{F}$ into its singleton components. Table~\ref{tab:mfvd} illustrates relative memory consumption for our solution in experiment one (Figure~\ref{fig:FMS09} with $N_Q=1000$).
	\begin{table}[htb]
	\renewcommand{\arraystretch}{1.0}
	\caption{Comparison in memory usage ($N_Q=1000$)}
	\label{tab:mfvd}
	\centering
		\begin{tabular}{|c|c|c|}
		\hline
		\textbf{Technique}		&		\textbf{\# of Counters}		& \textbf{SRAM usage}\\ \hline
		Per-Flow		&		8192 	&	100\%\\ \hline
		FlowME   	&		3555 	&    43\%\\ \hline
		\end{tabular}
	\end{table}
	
\subsection{Lattice structure generation time}

\begin{figure}[ht]
	     \centering
	     \includegraphics[scale=0.55]{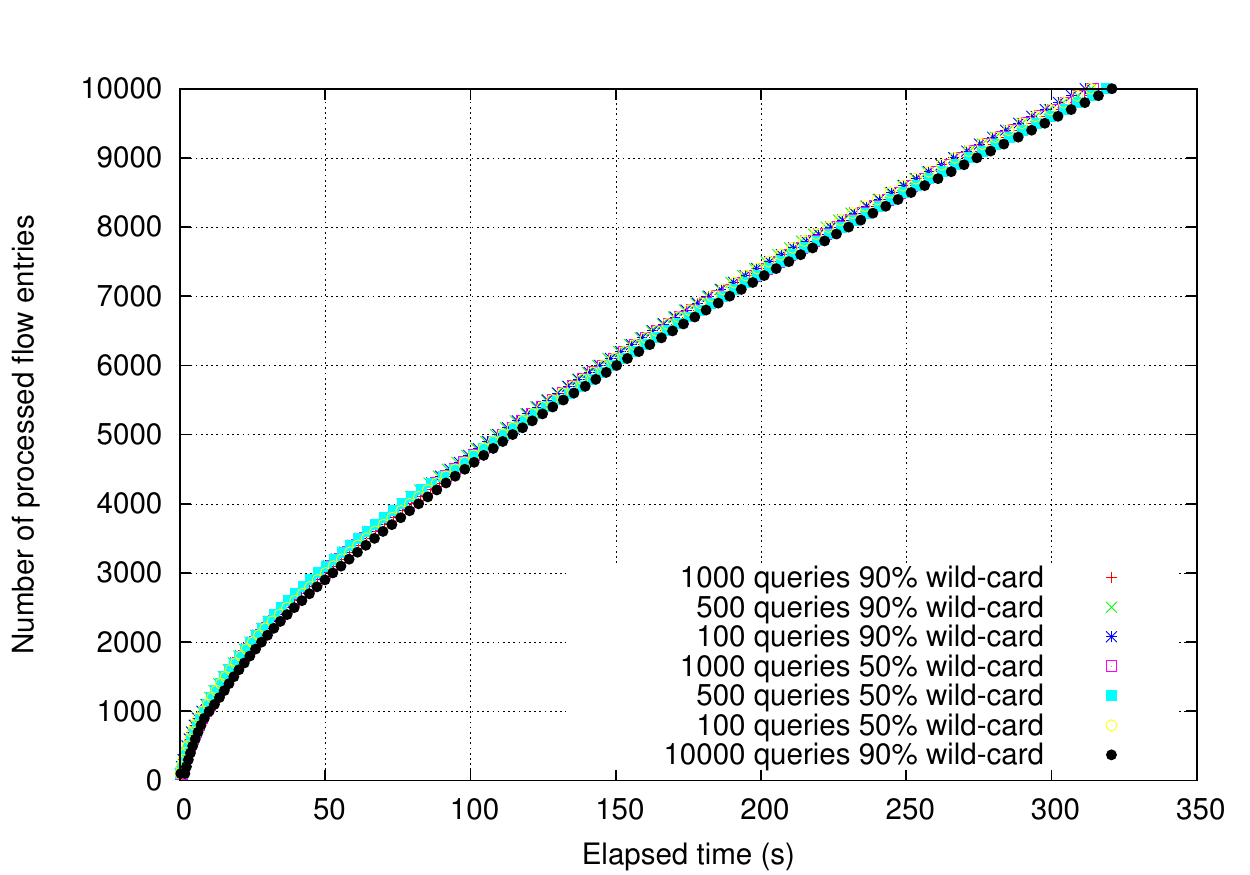}
		 \caption{Structure incrementation time: A group of 100 flow entries is added to the structure at each iteration}
	     \label{fig:SIT}
	 \end{figure}

The flexibility of FlowME is assessed by measuring the update time for a new flow entry. Two main operations are monitored: (1) lattice update and (2) ground concept identification and flow partition extraction. 
In our experiment, 10000 flow entries are split into groups of 100 to be sent for lattice update at subsequent steps of the incremental process. Figure~\ref{fig:SIT} shows the number of flow entries added at each step in the experimentation period. For instance, in the first 8s, 1000 flows are added. Overall, it took a total of 5min30s to build the lattice and identify flow entry partitions incrementally. 
	
\subsection{Packet processing performance}

The main challenge in traffic measurement is to minimize the impact on packet processing performance (time in clock cycles). In our testbed using a $400$MHz network processor, the average total packet processing time is $2447$ns. Since FlowME focuses only on flows covered by user queries, the remaining flows are automatically accelerated. As indicated above, we implemented a routine for the packet resolving stage that increments flow counters based on the reference retrieved at the searching stage. 
Processing time is now measured by placing start/end timestamps around the routine. According to the experimental outcome, 
the increment routine is performed in $9$ clock cycles. Hence, for an unmanaged flow packet, FlowME lowers the total processing time by~$22.5$ns. 

\section{Related work}
\label{sec:related-work}

The closest approach from the literature is the ProgME~\cite{Yuan:2011:PTP:1959441.1959451} traffic measurement tool. ProgME reflects application requirements through a rich query language where $\cap$, $\cup$ and $\setminus$ are used to compose queries from simpler ones.
To answer a set of queries, ProgME decomposes them into a set of disjoints flowsets, and assigns a counter to each one. In that, it does not rely on predefined flows. In contrast, our flows, and queries for that matters, are defined as 
simple conjunctions of matchfield values. However, the same set-theoretic operators on the answer sets of flows corresponding to queries are successfully simulated by our ground projection-based partition. In particular, the set of flows grounded in a projection represent the set-theoretic difference between the projection extent and the union of all extents of smaller projections. As a result, our own partition of $\mathcal F$ comprises a local partition of each answer set of flow entries. Now, our main advantage over ProgME lays in the (proven) minimality of counter-assignment solution: The lattice structure ensures none of the implicit set operations is redundant whereas the disentangling algorithm in ProgME lacks such a result.
%

AutoFocus~\cite{Estan:2003:AIP:863955.863972} is a tool for offline hierarchical traffic analysis whose goal is complementary to ours. Like ProgME, it doesn't use predefined flows but rather discovers them. To that end, it mines hierarchies of frequent generalized values for each matchfield and combines them into a global multidimensional structure. The structure comprises both the most significant and some deviant flows. As the authors themselves admit, the approach boils down to mining frequent generalized patterns on multiple dimensions. In comparison, our lattice contains the frequent \textit{closed} patterns of matchfield values from $\mathcal{F}$ which is a strict subset of all frequent patterns~\cite{dm-fca:valtchev+04}.
%
%

Current flow-based monitoring and collection systems like Cisco Netflow, FlowScan \cite{Plonka:2000:FNT:1045502.1045522} and sFlow~\cite{rfc3176} track all flow statistics continuously at a specific sampling rate. This generates a large number of transactions and a management bandwidth usage proportional to the number of flows, regardless of real application needs.
The comparison of FlowME to a flow-based measurement technique (section~\ref{sec:mem-cost}) shows the huge reduction in the number of managed flows.

Finally, since the introduction of metered traffic groups by the ISO accounting model \cite{ISO-7498} a number of architectures based on that notion were proposed in IETF internet RFCs (e.g.~\cite{rfc2722}). Identification of traffic groups remains an open problem and is typically solved by network operations personnel. FlowME is a significant step toward its automation.

For an in-depth coverage of the traffic measurement field readers are referred to~\cite{Yuan:2011:PTP:1959441.1959451}.

\section{Conclusion}
\label{sec:conclusion}
We presented FlowME, a lattice-based traffic measurement solution that, we believe, is a significant contribution to the field. Its mathematically founded approach amounts to partitioning the set of flow entries into a minimal number of subsets, each assigned a hardware counter. The main advantages thereof, efficiency in statistic computation and optimal resource usage, have been experimentally confirmed through an implementation over an OpenFlow switch: The results show both a significant reduction in the number of hardware counters (up to a factor of 10) and excellent performances. Moreover, our algorithmic methods are easy to implement while highly flexible and adaptable to a wide range of contexts.

Within a broader scope, a crucial benefit of our solution is its predictability: A preliminary assessment of the resources required by a user request is enabled in order to ensure their consumption respects the acceptable limits, i.e., 10\%. Overall, our approach will enable user control over the set of statistics in OpenFlow 1.3~\cite{openflow13} instead of fixing them within the standard. Furthermore, the genericity of the mathematical solution makes it particularly suitable to future network protocols with open sets of packet fields (CDN~\cite{1036038}, NDN~\cite{Jacobson:2009:NNC:1658939.1658941}, etc.). 
 
Finally, the high versatility of the lattice-based framework enables large variations in the problem settings. For instance, in the shorter term, we shall investigate the reduced substructures of the lattice as support for the counter assignment. 
%
%
%
%
%
%
%
%
%


\bibliographystyle{IEEEtran}
\bibliography{bib}

\end{document}